\documentclass{ws-procs975x65}

\usepackage{graphicx,epsfig, amsmath,amssymb,amsfonts}

\begin{document}

\title{TESTING LARGE-ANGLE DEVIATION FROM GAUSSIANITY IN CMB MAPS}

\author{A. BERNUI, M. J. REBOU\c{C}AS and A. F. F. TEIXEIRA}
\address{Centro Brasileiro de Pesquisas F\'{\i}sicas\\
Rua Dr.\ Xavier Sigaud 150, \  22290-180 Rio de Janeiro --
RJ, Brazil}

\vspace{-3mm}
\begin{abstract}
A detection of the level of non-Gaussianity in the CMB data is essential
to discriminate among inflationary models and also to test alternative
primordial scenarios. However, the extraction of primordial non-Gaussianity
is a difficult endeavor since several effects of non-primordial nature can
produce non-Gaussianity. On the other hand, different statistical tools
can in principle provide information about distinct forms of non-Gaussianity.
Thus, any single statistical estimator cannot be sensitive to all possible
forms of non-Gaussianity. In this context, to shed some light in the potential
sources of deviation from Gaussianity in CMB data it is important to use
different statistical indicators.
In a recent paper we proposed two new large-angle non-Gaussianity indicators
which provide measures of the departure from Gaussianity on large angular
scales. We used these indicators to carry out analyses of non-Gaussianity
of the bands and of the foreground-reduced WMAP maps with and without the \emph{KQ75}
mask. Here we briefly review the formulation of the non-Gaussianity indicators,
and discuss the analyses made by using our indicators.
\end{abstract}

\bodymatter
\vspace{-2mm}
\section{Introduction}\label{intro}

A detection of the level of primordial non-Gaussianity in the CMB
data is crucial to discriminate inflationary models and also to test
alternative scenarios for the physics of the early universe.
Clearly the study of detectable non-Gaussianities in the WMAP data ought to
take into account that they may have non-cosmological origins as, for
example, unsubtracted foreground contamination, unconsidered point sources
emission and systematic errors. 
Deviation from Gaussianity may also have a cosmic topology origin
(see, e.g., the review Refs.~\refcite{CosmTopReviews} and  related
Refs.~\refcite{TopDetec}).
If, on the one hand, different statistical tools can in principle provide
information about distinct forms of non-Gaussianity, on the other hand
one does not expect that a single statistical estimator can be sensitive
to all possible forms of non-Gaussianity in CMB data.
In view of this, a great deal of effort has recently gone into verifying the
existence of non-Gaussianity by employing several statistical
estimators (an incomplete list of references is given, e.g., in
Refs.~\refcite{Some_non-Gauss-refs,Bernui-Reboucas2009a,Bernui-Reboucas2009b}
and references therein).

Recently we have proposed\cite{Bernui-Reboucas2009a} two new large-angle
non-Gaussianity indicators, based on skewness and kurtosis of large-angle
patches of CMB maps, which provide measures of the departure from Gaussianity
on large angular scales.
We used these indicators to search for the large-angle deviation from
Gaussianity in both band and foreground-reduced maps\cite{Bernui-Reboucas2009b}
with and without a \emph{KQ75} mask (see also the related Refs.~\refcite{Related}).
Here we briefly summarize the main results of Refs.~\refcite{Bernui-Reboucas2009a}
and~\refcite{Bernui-Reboucas2009b}. 

\vspace{-2mm}
\section{Results and Concluding Remarks}

A constructive way of formulating our non-Gaussianity indicators $S$
and $K$  from CMB data is through the following steps:
\begin{alphlist}
\item
Take a finite set of points $\{j=1, \ldots ,N_{\rm c}\}$ homogeneously distributed
on the CMB  celestial sphere $S^2$ as the centers of spherical caps of a given
aperture $\gamma$; and calculate for each cap $j$ the skewness ($S_j$) and
kurtosis ($K_j$) by using that
\begin{equation}
S_j   \equiv  \frac{1}{N_{\rm p} \,\sigma^3_{\!j} } \sum_{i=1}^{N_{\rm p}}
\left(\, T_i\, - \overline{T_j} \,\right)^3
\quad \mbox{and} \quad
K_j   \equiv  \frac{1}{N_{\rm p} \,\sigma^4_{\!j} } \sum_{i=1}^{N_{\rm p}}
\left(\,  T_i\, - \overline{T_j} \,\right)^4 - 3 \,,
\end{equation}
where $N_{\rm p}$ is the number of pixels in the $j^{\,\rm{th}}$ cap,
$T_i$ is the temperature at the $i^{\,\rm{th}}$ pixel, $\overline{T_j}$ is
the CMB mean temperature of the $j^{\,\rm{th}}$ cap, and $\sigma$ is the
standard deviation.
The numbers $S_j$ and $K_j$ obtained in this way for each cap
can be seen as a measure of non-Gaussianity in the direction of
the center $(\theta_j, \phi_j)$ of the $j^{\,\rm{th}}$ cap.
\item
Patching together the $S_j$ and $K_j$ values for each spherical cap,
one obtains two discrete functions $S = S(\theta,\phi)$
and $K = K(\theta,\phi)$ defined on the sphere $S^2$, which can be
used as statistical indicators to measure the deviation from Gaussianity
as a function of the angular coordinates $(\theta,\phi)$.
The Mollweide projection of skewness and kurtosis
functions $S = S(\theta,\phi)$ and $K = K(\theta,\phi)$  are nothing but
skewness and kurtosis maps (hereafter $S-$map and $K-$map).
\end{alphlist}

Clearly, the discrete functions $S = S(\theta,\phi)$ and $K = K(\theta,\phi)$
can be expanded into their spherical harmonics in order to determine their power
spectra $S_{\ell}$ and $K_{\ell}$.
Thus, for example, for the skewness one has
$S (\theta,\phi) = \sum_{\ell=0}^\infty \sum_{m=-\ell}^{\ell}
b_{\ell m} \,Y_{\ell m} (\theta,\phi)$ and
$S_{\ell} = (2\ell+1)^{-1}\sum_m |b_{\ell m}|^2$. 
Similar expressions obviously  hold for the kurtosis $K = K(\theta,\phi)$.

In the remainder of this work we shall report the results of our Gaussianity
analysis performed with  $S = S(\theta,\phi)$ and $K = K(\theta,\phi)$
indicators calculated from single frequency and foreground reduced maps
with and without  a \emph{KQ75} mask.
To minimize the statistical noise, in the calculations
of $S-$map and $K-$map  from the
input maps, we have scanned the celestial
sphere with spherical caps of aperture  $\gamma = 90^{\circ}$,
centered at $12\,288$ points homogeneously distributed on the two-sphere.

For the sake of brevity we do note show examples of $S$ and $K$ maps,
which provide only \emph{qualitative} information on large-angle deviation
from Gaussianity (see figures in Refs.~\refcite{Some_non-Gauss-refs,%
Bernui-Reboucas2009a,Bernui-Reboucas2009b} for examples).
To obtain \emph{quantitative} information about the large angular
scale (low $\ell$) distributions for the non-Gaussianity $S$ and $K$ maps
obtained from the  CMB input maps used, we have calculated the (low $\ell$)
power spectra $S_{\ell}$ and $K_{\ell}$ for $S$ and $K$ maps. The deviation
from Gaussianity and the statistical significance were estimated
by comparing these power spectra with the corresponding averaged
power spectra $\overline{S}_{\ell}$ and $\overline{K}_{\ell}$ calculated
from $S$ and $K$ maps obtained by averaging over $1\,000$ Monte-Carlo-generated
statistically Gaussian CMB maps.
To have an overall assessment of low $\ell$ power spectra $S_\ell$ and $K_\ell$
calculated from each CMB input, we have performed a $\chi^2$
test to find out the goodness of fit for $S_{\ell}$ and $K_{\ell}$ multipole
values as compared to the expected multipole values from the MC Gaussian maps.
In this way, we obtained one number for each map that collectively ('globally')
quantifies the deviation from Gaussianity.
For the power spectra $S_\ell$ calculated from $S-$maps obtained from
the five-years maps with a \emph{KQ75} mask we found that the ratio
$\chi^2/\,\text{dof}\,$ (dof stands for degree of freedom) from the
K, Ka, Q, V, and W  maps  are given, respectively, by $21.5$, $4.9$,
$6.0$, $5.2$, and $3.9$,  while for the kurtosis power spectra $K_{\ell}$
of these maps the values of $\chi^2/\text{dof}$ are, respectively,
$35\,652$, $135$, $0.5$, $6.4$, and  $5.6$.
Clearly a good fit occurs when $\chi^2/\text{dof}\,\sim 1 $. Moreover,
the greater are the $\chi^2/\text{dof}\,$  values the smaller
are the $\chi^2$ probabilities, that is the probability that the multipole
values $S_{\ell}$ and $K_{\ell}$ and the expected MC multipole values agree.
For $S_{\ell}$ and $K_{\ell}$ obtained from the full-sky foreground-reduced
five years ILC input maps we found that $\chi^2/\text{dof}\,$ are,
respectively,  $35.7$  and $2\,368$. These results reduce to  $1.2$  and
$0.4$ when the \emph{KQ75} mask is employed. In brief, our analyses
show that the unmasked band maps are significantly non-Gaussian but
the deviation from Gaussianity is substantially reduced to a level
compatible with Gaussianity for Q and V maps, whereas the full-sky
foreground reduced five years ILC mask is  again significantly non-Gaussian
but the level of non-Gaussianity drops to a level that is consistent with Gaussianity
when the \emph{KQ75} mask is used.

\vspace{-3mm}
\section*{Acknowledgments}

A. Bernui and M.J. Rebou\c{c}as thank CNPq for the grants
under which this work was carried out. Some of the results
in this work were calculated by using the HEALPix~\cite{Gorski-et-al-2005}.

\end{document}